\documentclass[conference]{IEEEtran}
\IEEEoverridecommandlockouts
\usepackage{cite}
\usepackage{amsmath,amssymb,amsfonts}
\usepackage{algorithmic}
\usepackage{graphicx}
\usepackage{textcomp}
\usepackage{multirow,multicol}
\usepackage[flushleft]{threeparttable}

\usepackage{colortbl}
\usepackage{times}
\usepackage{pdfpages}
\usepackage{float}
\usepackage{afterpage}
\usepackage{svg}
\usepackage{array,ragged2e}

\newcolumntype{C}[1]{>{\centering\arraybackslash}p{#1}}

\usepackage{booktabs}
\usepackage{tikz}
\usepackage{pgfplots}
\usepackage{standalone}
\usepackage{titlesec}
\usepackage{enumerate}
\usepackage{soul}
\usepackage{svg}
\usepackage[utf8]{inputenc}
\usepackage{rotating}
\usepackage{afterpage}
\usepackage{multicol}
\usepackage{wrapfig} 
\usepackage{xcolor}
\definecolor{Emerald}{rgb}{0.31, 0.78, 0.47}
\colorlet{emerald_shape_1}{Emerald!20}
\colorlet{emerald_shape_2}{Emerald!40}
\colorlet{emerald_shape_3}{Emerald!60}
\colorlet{emerald_shape_4}{Emerald!80}
\colorlet{emerald_shape_5}{Emerald!100}
\colorlet{emerald_shape_6}{Emerald!120}
\colorlet{emerald_shape_7}{Emerald!140}

\usepackage{xurl} 
\usepackage{hyperref} 
\hypersetup{
 colorlinks=true,
 linkcolor=blue,
 filecolor=blue, 
 urlcolor=black,
 citecolor=blue
 }

\usepackage{enumitem,amssymb}
\newlist{todolist}{itemize}{2}
\setlist[todolist]{label=$\square$}
\usepackage{pifont}
\newcommand{\cmark}{\ding{51}}%
\newcommand{\xmark}{\ding{55}}%
\newcommand{\done}{\rlap{$\square$}{\raisebox{2pt}{\large\hspace{1pt}\cmark}}%
\hspace{-2.5pt}}
\newcommand{\wontfix}{\rlap{$\square$}{\large\hspace{1pt}\xmark}}

\usepackage{import}
\usepackage{longtable}
\usepackage{amsmath}
\usepackage{cleveref} 

\newboolean{showcomments}
\setboolean{showcomments}{true} 
\ifthenelse{\boolean{showcomments}}
{\newcommand{\nb}[2]{
 \fcolorbox{black}{yellow}{\bfseries\sffamily#1}
 {\sf$\blacktriangleright$\textit{#2}$\blacktriangleleft$}
 }
 
}
{\newcommand{\nb}[2]{}
 
}

\makeatletter
\def\ps@IEEEtitlepagestyle{%
  \def\@oddfoot{\mycopyrightnotice}%
  \def\@oddhead{\hbox{}\@IEEEheaderstyle\leftmark\hfil\thepage}\relax
  \def\@evenhead{\@IEEEheaderstyle\thepage\hfil\leftmark\hbox{}}\relax
  \def\@evenfoot{}%
}
\def\mycopyrightnotice{%
  \begin{minipage}{\textwidth}
  \centering \scriptsize
  This article has been accepted for publication in the 21st IEEE International Conference on Software Architecture (ICSA 2024). This is the author's version, which has not been fully edited, and content may change prior to final publication. 
  
  Copyright~\copyright~20xx IEEE. Personal use of this material is permitted.  Permission from IEEE must be obtained for all other uses, in any current or future media, including reprinting/republishing this material for advertising or promotional purposes, creating new collective works, for resale or redistribution to servers or lists, or reuse of any copyrighted component of this work in other works by sending a request to pubs-permissions@ieee.org.
  \end{minipage}
}
\makeatother

\begin{document}

\title{Architectural Design Decisions for Self-Serve Data Platforms in Data Meshes
}

\author{\IEEEauthorblockN{Tom van Eijk\IEEEauthorrefmark{1}}
\IEEEauthorblockA{\textit{JADS and Tilburg University} \\
's-Hertogenbosch, The Netherlands \\
t.m.h.vaneijk@tilburguniversity.edu} \\[0.1cm]
\and
\IEEEauthorblockN{Indika Kumara\IEEEauthorrefmark{1}}
\IEEEcompsocitemizethanks{\IEEEcompsocthanksitem\IEEEauthorrefmark{1}equal contribution}
\IEEEauthorblockA{\textit{JADS and Tilburg University} \\
's-Hertogenbosch, The Netherlands \\
i.p.k.weerasinghadewage@tilburguniversity.edu} \\[0.1cm]
\and
\IEEEauthorblockN{Dario Di Nucci}
\IEEEauthorblockA{\textit{University of Salerno} \\
Salerno, Italy \\
ddinucci@unisa.it} \\ [0.1cm]
\and
\IEEEauthorblockN{Damian Andrew Tamburri}
\IEEEauthorblockA{\textit{JADS and Eindhoven University of Technology} \\
's-Hertogenbosch, The Netherlands \\
d.a.tamburri@tue.nl} \\ [0.1cm]
\and
\IEEEauthorblockN{Willem-Jan van den Heuvel}
\IEEEauthorblockA{\textit{JADS and Tilburg University} \\
's-Hertogenbosch, The Netherlands \\
w.j.a.m.vdnheuvel@uvt.nl}
}

\maketitle
\IEEEpubidadjcol

\begin{abstract}
Data mesh is an emerging decentralized approach to managing and generating value from analytical enterprise data at scale. It shifts the ownership of the data to the business domains closest to the data, promotes sharing and managing data as autonomous products, and uses a federated and automated data governance model. The data mesh relies on a managed data platform that offers services to domain and governance teams to build, share, and manage data products efficiently. However, designing and implementing a self-serve data platform is challenging, and the platform engineers and architects must understand and choose the appropriate design options to ensure the platform will enhance the experience of domain and governance teams. For these reasons, this paper proposes a catalog of architectural design decisions and their corresponding decision options by systematically reviewing 43 industrial gray literature articles on self-serve data platforms in data mesh. Moreover, we used semi-structured interviews with six data engineering experts with data mesh experience to validate, refine, and extend the findings from the literature. Such a catalog of design decisions and options drawn from the state of practice shall aid practitioners in building data meshes while providing a baseline for further research on data mesh architectures.
\end{abstract}


\section{Introduction}
Companies are generating data at a rapid rate. The International Data Corporation predicts that global data will double in size between 2022 and 2026, and enterprise data will grow more than twice as fast as consumer data~\cite{IDC}. Despite the vast volumes of data companies have, the difficulties in integrating, managing, and governance them at scale have impeded utilizing them to unlock strategic insights and business growth~\cite{deloittedm}. In 2018, Zhamak Dehghani~\cite{dehghani2022data} proposed the data mesh approach for sharing, managing, and generating business value from analytical data in enterprises as a potential solution to addressing such issues. Since then, the data mesh has taken the attention of practitioners as evidenced by the growing number of gray literature on the topic~\cite{goedegebuure2023data} and the research based on industrial case studies~\cite{driessen2023promote,pakrashi2023cowmesh,driessen2022data,bode2023data,Eichler2023,migatraionstudy}. 

The data mesh approach is grounded in four principles: decentralized ownership of domain-specific data, a product-oriented mindset for analytical data, a federated data governance model, and a self-serve data platform for empowering domains to create high-quality data products with a high autonomy~\cite{dehghani2022data}. These principles aim to increase agility in value extraction from data by decentralizing data ownership to the business domains that produce data or have intimate knowledge of data and its usage by adopting product thinking to treat the data as interoperable products that meet the needs of data users and can be seamlessly combined to achieve a greater higher-order value. 

The key technical infrastructure of a data mesh is the self-serve data platform, which aims to lower the barriers for domain teams to own, build, and exchange data products and for governance teams to monitor and ensure interoperability, compliance, and quality of data products. When building a self-serve platform, a platform team (e.g., platform architects and engineers) needs to consider various design decisions and select the appropriate options for them to optimize the experience for data product teams, consumers, and governance teams~\cite{dehghani2022data,goedegebuure2023data,bode2023data}. While there are studies on identifying architectural design decisions (ADDs) in other domains, such as machine learning~\cite{Zdun9779697_icsa,Zdun9734278_ML} and blockchains~\cite{blockchain_adds}, there are no similar studies on data mesh. Thus, the work presented in this paper seeks to compile a catalog of ADDs and their options systematically, and the following central questions guide it:

\begin{center}
\textit{Which architectural design decisions (ADDs) can be chosen in the context of self-serve data platforms for data meshes? What are their options? How can a given option affect the experience of the data mesh stakeholders?}
\end{center}

We first performed a systematic review of the gray literature on self-serve data platforms to answer these questions, following the well-established guidelines on conducting such studies~\cite{garousi2019guidelines,kumara2021s}. As in similar studies~\cite{Zdun9734278_ML,Zdun9779697_icsa}, we selected the gray literature since we want to understand the design choices made by practitioners, and there is only a little academic literature on the theme. By applying appropriate coding methods~\cite{corbin2014basics}, we extracted the ADDs, their solution options, and their impact on the experience of the stakeholders in the data mesh. Next, we conducted semi-structured interviews with six participants with a strong data engineering background and at least one year's experience in data mesh. As in design science studies~\cite{oates2005researching}, we aimed to validate, refine, and enrich our framework created from the gray literature. Overall, we found six main ADDs and 55 options. We hope such findings will help organizations embark on their data mesh journey and research community to identify the design issues in the key technical infrastructure of the data mesh.

The paper is organized as follows. \Cref{sec:background} overviews data mesh, highlighting the key layers in the self-serve platform and reviewing the related studies. \Cref{sec:design} details the research methodology, encompassing systematic gray literature analysis and semi-structured expert interviews. \Cref{sec:framework} presents self-serve platform ADDs and options. \Cref{sec:validation} presents the expert feedback concerning our framework. \Cref{sec:threats} discusses  the threats to validity. Finally, \Cref{sec:conclusions} concludes the paper and suggests future work. 

\section{Background and Related Work}
\label{sec:background}
This section briefly explains the data mesh, highlights the layers of the self-serve platform, and reviews the related studies on data mesh. 

\subsection{Data Mesh and its Self-Serve Data Platform}
\label{sec-back-self-serve}
A data mesh is a socio-technical approach to sharing and managing analytical data in organizations~\cite{dehghani2022data} based on four principles: domain ownership of data, data-as-a-product, computational federated governance, and self-serve data platform. Business domains creating data assets (e.g., data, models, and dashboards) are responsible for managing assets and offering them as products that create value for some consumers. A data product ingests data from data sources and other data products (via its input interfaces/ports), transforms the ingested data to the results expected by the data product consumers, and makes the results available to the consumers (via its output interfaces/ports). Data products are the units of value exchange in a data mesh and must feature several non-functional properties, including discoverable, interoperable, and valuable~\cite{dehghani2022data,goedegebuure2023data}. Federated computational governance ensures adherence to global interoperability standards and policies, facilitating value extraction from aggregated and correlated independent data products. Self-serve platforms empower cross-functional domain teams to autonomously build, share, manage, and consume data products and the governance team to ensure the regulatory and policy compliance of the data mesh. 


\begin{figure}[ht]
 \centering
 \includegraphics[width=0.55\columnwidth]{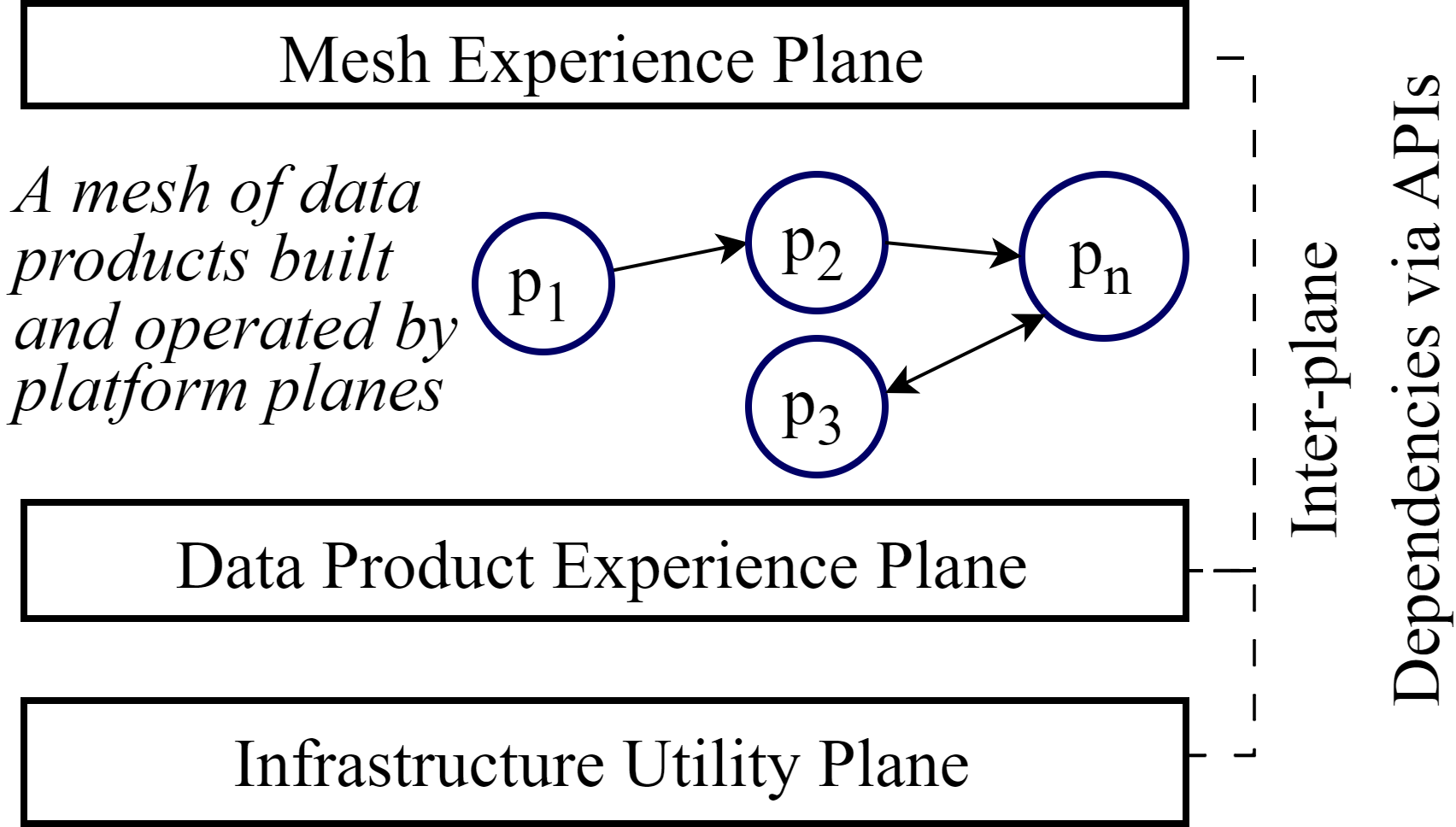}
 \caption{Multiple planes of a self-serve data platform~\cite{dehghani2022data}}
 \label{fig:self-serve-multi-planes}
\end{figure}


This paper focuses on the self-serve platform, which provides the foundation for building and governing a mesh of data products. \Cref{fig:self-serve-multi-planes} shows the logical architecture of a self-serve platform, which consists of three layers: data infrastructure utility plane, data product experience plane, and mesh experience plane~\cite{dehghani2022data}. The infrastructure plane offers the infrastructure and platform resources (e.g., VMs, storage, networking, data transformation engines, and policy engines) required for operating the data mesh. The data product experience plane aims to hide the complexities of using the infrastructure plane in creating, maintaining, and consuming individual data products. Data product developers benefit from functionalities such as building, deploying, and testing data products, while consumers can easily access and utilize data products for their specific needs. Finally, the mesh experience plane operates at the overarching mesh level, abstracting capabilities that involve multiple data products. It acts as a cohesive layer aggregating interfaces from the data product experience plane. It provides a holistic view of the mesh, enabling a comprehensive and interconnected experience for users navigating the entire data mesh.

\subsection{Related Work}
The main goal of this study is to understand the architectural design decisions that a platform team needs to make when designing a self-serve platform. While recent works exist on data mesh, to our knowledge, there is no study on architectural design decisions and their options related to the data mesh. 

Goedegebuure et al.~\cite{goedegebuure2023data} conducted a systematic review of the gray literature on data mesh. They analyzed the practitioner's perspective on four principles of data mesh and created three reference architectures for data mesh. Bode et al.~\cite{bode2023data} conducted semi-structured interviews with practitioners to understand the organizational motivation for adopting data mesh, potential challenges, and implementation strategies. Machado et al.~\cite{MACHADO2022263} reviewed the academic literature on data mesh and analyzed two implementations (mentioned in the gray literature). However, they did not identify design decisions or options. 
Loukiala et al.~\cite{migatraionstudy} presented the lessons from a data mesh migration study in a large manufacturing country. They used the tools a public cloud provider offers to implement a self-serve platform without providing information concerning their design decisions. 

Falconi and Plebani~\cite{falconi2023adopting} attempted to adopt a data mesh architecture for sharing data among federated organizations in clinical trials. In particular, they proposed a new logical plane for the self-serve platform, namely the federated data mesh experience plane, to ensure continuous alignment of information about the federation's status among participating organizations. The data platform implementation by Wider et al.~\cite{wider2023decentralized} revolves around data products interacting with their environment, i.e., the mesh, through ports. Key components include an operator app, a data catalog, and a governance layer. The operator app orchestrates tools in the ecosystem, managing the lifecycle of data products. The data catalog enables data discoverability and lineage, while the governance layer consists of tools managing governance rules, such as access control at both the mesh and data product levels.
Ashraf et al.~\cite{ashraf2023key} evaluated the tools supporting the data mesh principles for a data platform, categorizing them based on use cases, including metadata management, deployment, standardizing data product APIs, and data observability. Eichler et al.~\cite{Eichler2023} designed and implemented a data marketplace for sharing data in an enterprise. Their implementation uses a central data catalog to store metadata about data assets.
 
The existing studies have analyzed some implementations of data mesh, implemented several components of a self-serve platform, and created reference architectures for data mesh. However, none have systematically formulated the guidelines for designing a self-serve data platform. 

\section{Research Design}
\label{sec:design}

In this paper, we conducted a systematic gray literature review (SGLR) concerning self-serve data platforms in data meshes to understand design decisions made by practitioners. We applied the coding methods from the grounded theory ~\cite{corbin1990grounded} to the gray sources to identify the design decisions and options and interviewed experts to validate and extend them.

\subsection{Design Design Framework Creation}
We applied the guidelines and process proposed by Garousi et al.~\cite{garousi2019guidelines} and other gray literature studies (e.g.,~\cite{kumara2021s}) to conduct the review of the literature on self-serve platforms. We chose the gray literature as the sole data source for two reasons. On the one hand, we aimed to understand the design decisions made by the practitioners, as in similar studies~\cite{Zdun9779697_icsa,Zdun9734278_ML}. On the other hand, only a limited academic literature exists on the data mesh domain. 

We focus solely on textual sources like blog posts, white papers, slide decks, and reports. We searched the initial sources using Google Search. Following Garousi et al.'s recommendations~\cite{garousi2019guidelines}, we diligently explored reference lists and backlinks from the initial sources to identify more gray literature. We then applied the criteria for including and excluding sources and assessing source quality to refine the relevant literature corpus. We use the standard criteria used by other gray literature reviews~\cite{kumara2021s,garousi2019guidelines}; for example, we filtered articles according to the focus of the study (i.e., architectural decisions, options, and drivers), selecting papers written in English with reputable publishers and expert authors. The sources not meeting the defined quality standards were systematically excluded from the review. The first two authors of the paper independently selected the studies by searching in incognito mode to avoid personal biases. The inter-rater reliability (Cohen's Kappa) was 0.64, indicating substantial agreement. After an iterative screening of sources, 43 articles met the qualitative assessment criteria and were included in the final study.

\begin{table}[ht]
\centering
\caption{Results of the Search Queries for ``Self-Serve Platform''}
\resizebox{0.99\linewidth}{!}{
\begin{tabular}{|l|c|c|c|c|} 
\hline
\textbf{Search Query} & \textbf{Initial Results} & \textbf{1st Screening} & \textbf{2nd Screening} & \textbf{QAC} \\ 
\hline
Self-Serve Data Platform/Infrastructure & 133 & 32 & 18 & 17 \\ 
\hline
Self-Service Platform/Infrastructure & 218 & 25 & 16 & 13 \\ 
\hline
Data Mesh Platform/Infrastructure & 116 & 11 & 8 & 7 \\ 
\hline
Data Mesh Ecosystem & 55 & 13 & 10 & 6 \\ 
\hline
\textbf{Total} & \textbf{522} & \textbf{81} & \textbf{52} & \textbf{43} \\
\hline
\end{tabular}
}
\label{table:query-result}
\end{table}

\Cref{table:query-result} summarizes the search queries and the number of articles found and included.
We analyzed the selected sources using qualitative data generation techniques from grounded theory (GT)~\cite{corbin1990grounded}. As in similar studies~\cite{Zdun9779697_icsa,Zdun9734278_ML}, we employed three coding activities: open coding for developing categories, axial coding for refining and linking categories, and selective coding for synthesizing categories into core concepts~\cite{corbin1990grounded}. Coding continued until theoretical saturation, where no new information on properties, dimensions, or relationships emerged~~\cite{corbin2014basics}. We used memo writing to document the theory-building process and enhance transparency. In particular, we used Markdown files to preserve open and axial coding, ensure comprehensive audit trails, and promote research reproducibility and traceability of codes to their sources. For architectural decision modeling, we used CodeableModels~\cite{Zdun9734278_ML}, a Python package for defining models and instances through code and memos. The first author of this paper coded all sources, and the second author reviewed the generated codes. They resolved all the discrepancies via discussions.

\subsection{Interview-based Refinement and Validation}
Engaging with experts can provide valuable insights, deepening our understanding of practitioners' perspectives. Hence, we conducted semi-structured interviews to validate and improve our findings and ensure the reliability of the results. An ex-ante evaluation with domain experts is generally used in design science research~\cite{oates2005researching}. 
We applied \textit{purpose sampling}~\cite{etikan2016comparison} to select participants with relevant expertise and varied viewpoints aligned with the data mesh paradigm. Candidates must have at least three years of experience in data engineering and one year of experience in data mesh. To find potential candidates, we searched in a data mesh community\footnote{https://datameshlearning.com/}, LinkedIn, and the companies a data mesh research group recommended.

\begin{table}[ht]
 \caption{Interviewed Experts}
 \label{tab:interview-participants}
 \centering
 \begin{threeparttable}
\resizebox{0.99\linewidth}{!}{
\begin{tabular}{|l|l|l|r|c|r|r|r|r|} 
\hline
\textbf{ID} & \textbf{Type of stakeholder} & \textbf{Industry} & \begin{tabular}[c]{@{}l@{}}\textbf{Company }\\\textbf{Size (\#emps.)}\end{tabular} & \textbf{CC} & \textbf{\#O} & \textbf{\#DE} & \textbf{\#DM} & \textbf{Length} \\ 
\hline
p1 & Intel. Platform Lead & FinTech & 30,000 & FR & 24 & 28 & 1.5 & 26:22 \\ 
\hline
p2 & Data Arch. \& Proj. Lead & Software & 29 & IT & 2 & 5 & 1.0 & 57:02 \\ 
\hline
p3 & Data Mesh Comm. Lead & Media & 5,500 & BE & 4 & 15 & 1.5 & 54:48 \\ 
\hline
p4 & Exec. IT Consultant & FinTech & 298,000 & CA & 17 & 38 & 1.5 & 56:02 \\ 
\hline
p5 & Data Mesh Manager & FinTech & 35,000 & US & 4 & 3 & 1.0 & 54:01 \\ 
\hline
p6 & Data Eng. Manager & e-Commerce & 400 & ES & 9 & 16 & 2.0 & 46:57 \\
\hline
\end{tabular}
 }
 \begin{tablenotes}
 \item \textbf{ID}: Pseudonym for the expert
 \item \textbf{CC}: Expert's country of origin (ISO Alpha-2 codes)
 \item \textbf{\#O}: Number of organizations where the expert has worked
 \item \textbf{\#DE}: Years of experience in data engineering/management
 \item \textbf{\#DM}: Years of experience in data mesh practices
\end{tablenotes}
 \end{threeparttable}
 \end{table}

\begin{table}[ht]
\centering
\caption{Evaluation Criteria}
\label{tab:SelfServeOperationalisedEvaluationCharacteristics}
\resizebox{0.99\linewidth}{!}{
 \begin{tabular}{|p{1.6cm}|p{5cm}|p{5cm}|} 
\hline
\textbf{Characteristic} & \textbf{Definitions by \cite{davis1989perceived} and \cite{prat2015taxonomy}} & \textbf{Adapted Operational Definitions~} \\ 
\hline
Completeness & The extent to which the structure incorporates essential components and interrelationships. & Whether the self-serve platform framework contains all required decisions, decision options, and forces.\\ 
\hline
Perceived Usefulness & The extent to which using the framework enhances job performance & Whether the framework can provide guidance during~ designing a self-serve platform.\\ 
\hline
Perceived Ease of Use & The extent to which using the framework is free of effort & Whether the framework is easy to understand in practice.\\
\hline
\end{tabular}
 }
\end{table}

\Cref{tab:interview-participants} provides the information about the six interviewed experts. The average interview duration is approximately $50.70$ minutes, with a standard deviation of approximately $10.34$ minutes.
The questions of the interviews were drawn from the technology acceptance model~\cite{davis1989perceived} and evaluation criteria hierarchy~\cite{prat2015taxonomy}. In particular, we considered \textit{Perceived Usefulness}, \textit{Perceived Ease of Use}, and \textit{Completeness}, as defined in \Cref{tab:SelfServeOperationalisedEvaluationCharacteristics}. 
We created an interview protocol aligned with the interview protocol refinement framework~\cite{castillo2016preparing} to clarify research objectives, interview structure, confidentiality, and recording permission. We communicated the protocol to the experts before the interviews, which we conducted in a face-to-face online environment. After the interviews, we performed an audit trail, a crucial validity measure, by providing interview transcripts to participants for review and validation~\cite{reza2017comprehensive}. These interviews, lasting $45$ to $60$ minutes, were transcribed within $24$ hours. 
We systematically performed thematic analysis~\cite{clarke2015thematic} on the transcripts, aligning with our gray literature analysis.

\subsection{Replication Package}

The replication package of our study is available online\footnote{https://tinyurl.com/2d44n8c4}. It includes the complete list of sources, source selection criteria, quality assessment criteria, coded literature and interview transcripts, the interview guide, and the ethical approval for conducting the interviews.

\begin{table*}
\centering
\caption{Overview of Self-Serve Platform Architectural Design Decisions}
\resizebox{\textwidth}{!}{
\scriptsize
\begin{tabular}{|p{3cm}|c|p{4cm}|p{9cm}|p{3cm}|}
\hline
\textbf{Design Decision} & \textbf{\#} & \textbf{Concrete Decision Options} & \textbf{Evidence} & \textbf{Drivers/Forces} \\ 
\hline
\multirow{15}{*}{\parbox{3cm}{Which capabilities/APIs should be offered by the infrastructure utility plane for executing data product components, and how?}} & 23 & Batch/Stream~ Data Source/Sink Connectors & s3, s4, s12-s14, s18, s20, s22, s25, s27, s28, s30, s31-s34, s40, s41, p1-p6 & FC+, FD+, FP+ \\ 
\cline{2-5}
 & 14 & Legacy and Operational System Connectors & s1, s7, s8, s14, s16, s27, s30-s33, s40, s42, p1, p3 & FC+, FD+, FP+ \\ 
\cline{2-5}
 & 7 & BI Tools Connectors & s6, s10, S23, s31-s33, s41 & FC+, FD+, FP+, FG+ \\ 
\cline{2-5}
 & 12 & Federated Query Engine & s5, s27, s28, s30, s32-s34, s38, s41, s43, p3, p4 & FC+ \\ 
\cline{2-5}
 & 8 & Connector Repository & s2, s4, s6, s28, s32, s38, s40, p2 & FC+, FD+, FP+ \\ 
\cline{2-5}
 & 27 & Batch/Stream Data Processors & s5, s6, s11, s12, s13, s16, s17, s18, s26, s27, s28, s30, s31, s33, s35, s37, s38, s40, s41, s42, s43, p1-p6 & FC(+/-), FD(+/-) \\ 
\cline{2-5}
 & 19 & BI Tools & s6, s10, s16, s17, 18, s20, s23, s25, s30-s32, s41, s42, p1-p6 & FC+, FD+, FG+ \\ 
\cline{2-5}
 & 20 & Polyglot Data Storage & s1, s3, s5-s7, s11, s13, s15, s16, s18, s26, s27, s29, s30-s35, s40, s41, s43, p3 & FC+, FD+, FP- \\ 
\cline{2-5}
 & 13 & Event Streaming Platform & s2, s6, s10, s17, s18, s32, s33, p1-p6 & FC+, FD+ \\ 
\cline{2-5}
 & 12 & Schema Registry & s4, s8, s12, s16, s32, s35, s36, p1-p5 & FC+, FD+ \\ 
\cline{2-5}
 & 13 & Metadata Store & s6, s9, s22, s31, s44, s40, s43, p1-p6 & FC+, FD+ \\ 
\cline{2-5}
 & 9 & Model Store & s18, s33, s38, p1-p6 & FC+, FD+ \\ 
\cline{2-5}
 & 14 & Data and ML Pipeline & s9, s16, s10, s18, s19, s21, s26, s31-s33, s38, s40, p2, p5 & FD+ \\ 
\cline{2-5}
 & 3 & Pipeline Connectors & s31, s22, p4 & FD+, FP+ \\ 
\cline{2-5}
 & 5 & Pipeline Templates & s17, s20, s25, s31, s42 & FD+ \\ 
\hline
\multirow{11}{*}{\parbox{3cm}{Which capabilities should be offered by the infrastructure utility plane for various governance functions at the product level and the mesh level, and how?}} & 28 & Data Catalog & s1, s3, s9, s10, s13-s18, s21, s24, s27-s30, s32,s33, s35-s37, s38, s42, s43, p1-p6 & FC+, FD+, FG+ \\ 
\cline{2-5}
 & 28 & API Catalog & s1, s7, s8, s9, s10, s11, s14, s15, s17, s18, s21, s25, s29, s30-s32, s34, s37-s42, p1-p6 & FC+, FD+ \\ 
\cline{2-5}
 & 8 & Pull/push Catalog Loading & s18, s10, s17, s21, s30, s40, s43, p2 & FD+, FG+ \\ 
\cline{2-5}
 & 3 & Pull/push Change Propagation & s4, s32, p2 & FC+, FD+, FG+ \\ 
\cline{2-5}
 & 13 & Policy-as-Code Tools & s5, s9, s11, s17, s29, s30, s34. s36, s40, p2, p4, p5, p6 & FD+, FG+ \\ 
\cline{2-5}
 & 12 & Data Quality Checkers & s9, s17, s24, s32, s33, s36, p1-p6 & FC+, FD+, FG+ \\ 
\cline{2-5}
 & 26 & Access and Identity Manager & s2, s3, s5, s8, s9, s11, s13, s15-s17, s26-s29, s31-s33, s33, s35, s42, s43, p1-p6 & FD+, FC+, FG+, FP(+/-) \\ 
\cline{2-5}
 & 6 & Privacy-Enhancing Technologies & s17, s21, s26, s31, s34, s40 & FC+, FD(+/-), FG+, FP- \\ 
\cline{2-5}
 & 9 & Resource Usage and Cost Monitoring & s9, s16, s17, s18, s30, s31, s42, p2, p5 & FC+, FD+, FP+, FG+ \\ 
\cline{2-5}
 & 7 & Log Management & s9, s17, s18, s31, p3, p4, p7 & FD+, FP-, FG+ \\ 
\cline{2-5}
 & 4 & Alert Generation & s6, s17, s42, p6 & FC+, FD+, FP+, FG+ \\ 
\hline
\multirow{8}{*}{\parbox{3cm}{Which capabilities should be offered by the infrastructure plane for deploying products, and how?}} & 5 & VMs & s14, s26, s30, s33, p4 & FP(+/-) \\ 
\cline{2-5}
 & 6 & Containers & s10, s26, s29, s30, p2, p5 & FD+,~FP(+/-) \\ 
\cline{2-5}
 & 3 & FaaS & s3, s30, s32 & FP(+/-) \\ 
\cline{2-5}
 & 7 & Networking & s1, s11, s14, s17, s31, s33, p6 & FC+, FD+, FP(+/-) \\ 
\cline{2-5}
 & 2 & Build Scripts and UI & s34, p5 & FD-, FP- \\ 
\cline{2-5}
 & 9 & Infrastructure as Code & s2, s4, s6, s17, s18, s33, s38, s39, p5 & FD+, FP(+/-) \\ 
\cline{2-5}
 & 11 & Multi-tenancy & s1, s2, s6,s8,s26,s42, s22, p2, p3, p5, p6 & FP(+/-) \\ 
\cline{2-5}
 & 2 & Virtual Private Cloud & s22, s26 & FD+, FP(+/-) \\ 
\hline
\multirow{4}{*}{\parbox{3cm}{What are~the cross-cutting capabilities/processes of the infrastructure plane?}} & 15 & CI-CD Pipeline Automation & s4, s10, s12, s17, s22, s26, s29, s33, s39, p1-p6 & FD+, FP(+/-) \\ 
\cline{2-5}
 & 14 & Plane APIs & s5, s10, s9, s11, s22, s30, s31, s33, s34, s39, p4-p6 & FD(+/-), FP(+/-), FG(+/-) \\ 
\cline{2-5}
 & 5 & Buy or Buy Component & s17, s16, s22, s33, s40 & FC(+/-), FD(+/-), FP(+/-), FG(+/-)\\ 
\cline{2-5}
 & 11 & Personas-first or Tool-first & s4, s6, s11, s23, s26, s27, s31, s33, s36, s40, p1 & FC(+/-), FD(+/-), FP(+/-), FG(+/-) \\ 
\hline
\multirow{12}{*}{\parbox{3cm}{Which capabilities should be offered by the product experience plane, and how?}} & 22 & Data Product Lifecycle APIs & s3, s5, s6, s9, s11, s13-s18, s25, s31, s33, s39, s42, p1-p6 & FD+, FG+ \\ 
\cline{2-5}
 & 10 & Version Data Product & s5, s30, s31, s33, p1-p6 & FC+, FD+, FG+ \\ 
\cline{2-5}
 & 25 & Govern Data Product & s2, s3, s5, s7, s8, s9, s11, s13, s15-s17, s24, s26-s33, s35, s42, s43, p1, p3, p4 & FC+, FD+, FG+, FP(+/-) \\ 
\cline{2-5}
 & 16 & Connect and Read Data Product & s2, s5, s9, s10, s11, s26, s27, s30, s33, s42, p1-p6 & FC+,FD+ \\ 
\cline{2-5}
 & 17 & Product Blueprints/Templates & s4, s5, s11-s13, s17, s19, s20,s27, s28, s30, s31, s33, s39, p3, p4, p6 & FD+ \\ 
\cline{2-5}
 & 5 & IaC Blueprints/Templates & s2, s30, s33, s39, p5 & FD+,FP+ \\ 
\cline{2-5}
 & 7 & Data Product Contract & s11, s17, s34, s40, p1, p2, p5 & FC+, FD+, FG+ \\ 
\cline{2-5}
 & 1 & Data Product Code Repository & p4 & FC+, FD+ \\ 
\cline{2-5}
 & 3 & Data Product Metrics & s17, s24, p4 & FC+, FG+ \\ 
\cline{2-5}
 & 2 & Metric Calculation Pipeline & s17, p4 & FD+ \\ 
\cline{2-5}
 & 7 & Feedback Loop & s10, s21, s30, s39, p3, p4, p5 & FC+, FD+ \\ 
\cline{2-5}
 & 3 & Push/Pull Propagation of Product Changes & s4, s17, s32, p2 & FC+, FG+ \\ 
\hline
\multirow{5}{*}{\parbox{3cm}{What capabilities should be offered by a data mesh experience plane, and how?}} & 16 & Register and Search Data Products & s5, s14, s17, s2, s27, s31, s33, s35, s37, s38, p1-p6 & FC+, FD+, FG+ \\ 
\cline{2-5}
 & 7 & Data Product Registry/Catalog & s8, s27, s33, s34, s37, s38, p5 & FC+, FD+, FG+ \\ 
\cline{2-5}
 & 14 & Data Mesh Dashboard & s8, s11, s16, s17, s21, s24, s31, s35, p1-p6 & FC+, FD+, FG+ \\ 
\cline{2-5}
 & 15 & Global Policy Enforcement & s3, s5, s15, s17, s21, s25, s33, s26, s30, p1-p6 & FC+, FD+, FG+ \\ 
\cline{2-5}
 & 5 & Data Product Contract Enforcement & s17, s3, s31, s32, p1 & FC+, FD+, FG+ \\ 
\hline
\multicolumn{5}{|p{1.2\textwidth}|}{FD: Product Developer Experience, FC: Product Consumer Experience, FP: Platform Team Experience, FG: Governance Team Experience, +: Positive, -: Negative, +/-: Positive or Negative depending on profiles of the team, the buy/build decision, or other option selections} \\
\hline
\end{tabular}
}
\label{OverviewTableSelfServe}
\end{table*}

\section{A Design Decision Framework for Self-Serve Data Platform}
\label{sec:framework}
This section presents Architectural Design Decisions (ADDs) derived from analyzing the gray literature and the interview data. \Cref{OverviewTableSelfServe} shows ADDs, the number of their associated sources, their concrete decision options, the evidence supporting them (gray literature denoted as '$s$' and interview participants as '$p$'), and their drivers. The expert interviews helped not only validate the results from the gray literature but also refine it, including identifying new options. Hence, the table includes the consolidated and refined findings from both studies. The options are not mutually exclusive. For the drivers, we only consider the impact on the experience of different stakeholders in a data mesh, as the objective of the self-serve platform is to optimize stakeholders' experience in developing, consuming, and governing data products~\cite{dehghani2022data}. The ADDs are mapped to the three planes of a self-serve platform (see \Cref{sec-back-self-serve}). We manually draw the diagrams based on the models generated by \textit{CodeableModels}. We significantly simplified the diagrams to improve their readability for space limitations. We shall not claim any research contributions from using a specific notation.


\subsection{Decisions concerning the Data Infrastructure Utility Plane}
The decisions related to the infrastructure plane concern: Product Component APIs, Governance Support APIs, Deployment APIs, and Crosscutting Decisions. 

\begin{figure*}[ht]
 \centering
 \includegraphics[width=0.90\textwidth]{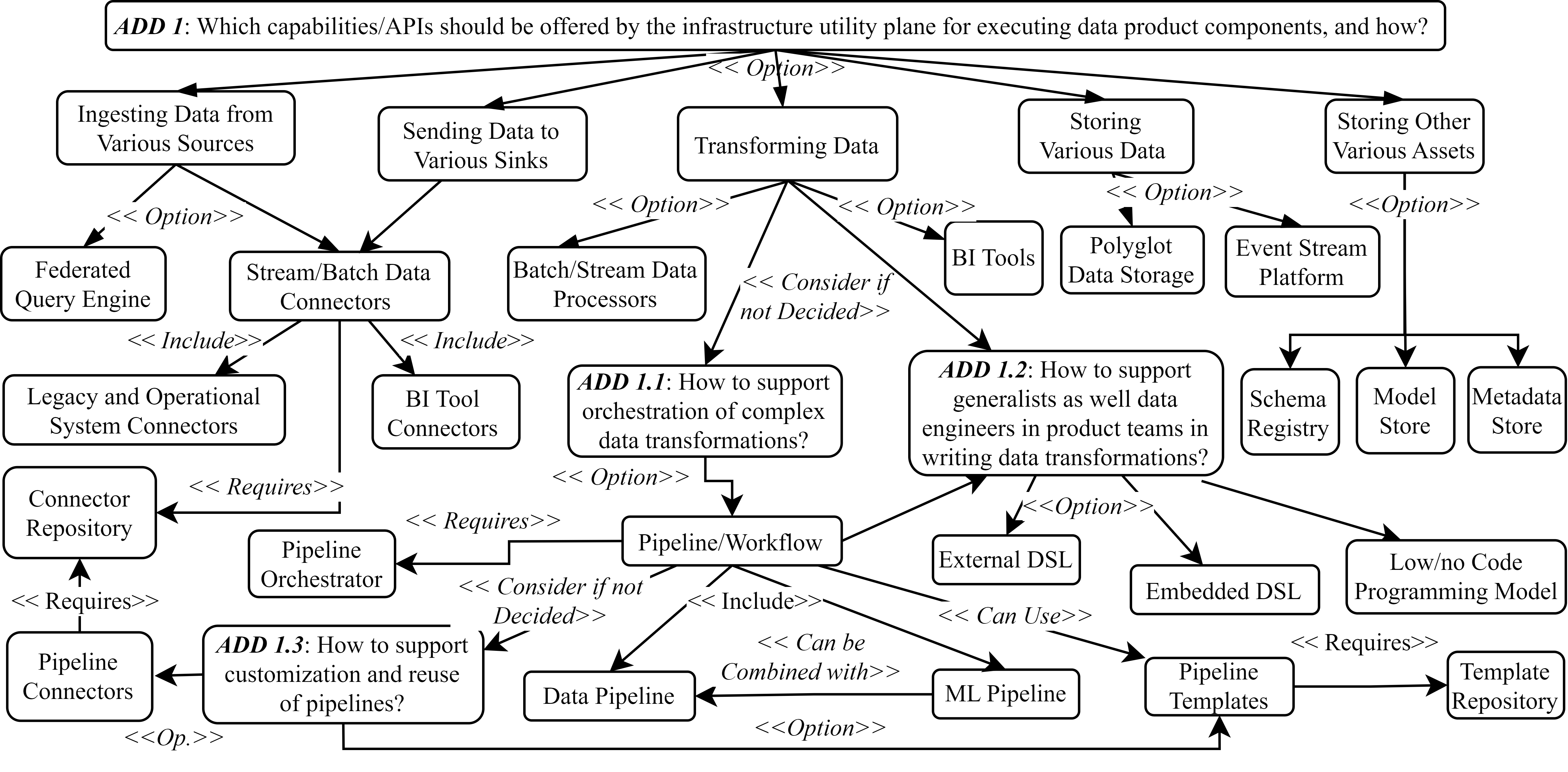}
 \caption{Decisions concerning Product Component APIs}
 \label{fig:infra_product_component}
\end{figure*}

\textbf{Product Component APIs.}
\Cref{fig:infra_product_component} shows the design decisions and their options to consider when designing the API of a data infrastructure plane to support the development and execution of product components. 

A data product uses a data ingestion task to consume the data it requires from data sources and other data products (i.e., the input ports of a data product). It may need to continuously read data from a data source (e.g., a topic in a streaming platform) or query it from a data source (e.g., a file in a cloud storage bucket) on demand. Hence, the infrastructure plane needs to include the components that can connect to and read from different data sources. According to our sources, the data in legacy and operational systems (e.g., ERP systems) can provide valuable insights for businesses; thus, the platform needs to provide the connectors to those systems. Furthermore, a data product often needs to combine data products from multiple domains efficiently. Therefore, the platform must also offer a federated query engine to execute queries to access data from different sources (e.g., object storage, relational databases, and APIs). 

A data product integrates and transforms the data received from the sources to produce what the product consumers expect. A data transformation may be triggered on demand, on a schedule (i.e., batch data processing or ad-hoc one-time query), or as data arrive (i.e., continuous stream processing). The infrastructure plane should host the appropriate data processing engines and provide an API to run the transformation code using them (e.g., to submit a Spark job to a managed Spark cluster). The specific types of data processing mentioned by our sources are ETL (Extract Transform Load) and ELT (Extract Load and Transform) pipelines, training a machine learning model, making inferences from such a model, and BI (Business Intelligence) reporting and dashboarding. By supporting common types of data transformations, the platform can enable the development of different types of data products. 

Data products can produce data assets (mainly data and models) in various formats and modes, e.g., an ML(machine-learning) model as a pickle file, a feature set as a file in a storage bucket or a table in a database, and the inferences from a model as an event stream. Consequently, the infrastructure plane should provide the capabilities to serve and share the produced artifacts using the access modes the products want to offer. For example, an event streaming platform such as Kafka can be hosted, and the connectors can be offered to submit data to it from the data transformation engines. A key decision option highlighted by the practitioners is to use polyglot storage that integrates multiple types of data storage systems, mainly relational databases, NoSQL(not only SQL), and object stores, and provides a unified abstraction to access/store the same data in different formats. A data product should be able to attract multiple consumers and thus should make it easy to consume the data. The data consumers should be able to use the data from a data product with little to no adaptation efforts and, thus, should be able to use the data in the most convenient format. 

Platform teams should be able to create, share, and update data source/sink connectors, and domain teams should be able to review and deploy them. Moreover, some reusable connectors may be harvested from the custom connectors implemented by domain teams. Our sources indicate that a common solution for addressing these issues is using a connector repository and providing the mechanisms to populate it and install connectors from it easily.   

A data product may produce or use specific artifacts, e.g., models, features, schemas, and metadata. Thus, the platform team must decide whether to use general-purpose storage, e.g., an object store, or custom storage, e.g., as a model store, feature store, schema registry, or metadata store. These custom storage options typically provide versioning, access control, and lineage tracking features, benefiting developers and consumers. Adding such features to general-purpose storage can be challenging and time-consuming, limiting the platform team's capability to serve the needs of the domain teams. 

A complex data transformation job may involve multiple steps, each consuming and producing some data assets, and a common practice to improve productivity in creating and executing such data transformation jobs is to develop them as workflows or pipelines, an ordered set of activities that consume and produce data assets. The pipelines enable the automated execution of the data transformation jobs. Hence, an infrastructure plane should include pipeline engines (or orchestrators) that support the development and execution of the common types of pipelines, such as data (engineering) pipelines, model training pipelines, and inference pipelines. Pipeline orchestrators are diverse in their programming models (i.e., imperative task-driven and event-driven) and languages (i.e., low/no-code visual languages, domain-specific languages, and general purpose). In the data mesh context, the selection of an orchestrator needs to consider how easy it is for the product developers from various domains to learn and adapt it, in addition to the common factors such as cost, performance, and scalability. The options for improving the product developer experience when using orchestrators include making the pipelines configurable by parameterization, a repository of reusable pipeline templates, and a repository of pipeline connectors/components. Custom pipeline connectors can allow product developers to easily interact with various external systems (e.g., cloud providers or legacy systems) and execute specific tasks (e.g., sending an email, masking sensitive data, or storing data in a cloud storage service), simplifying the development of complex data products. 

\begin{figure*}[ht]
 \centering
 \includegraphics[width=0.99\textwidth]{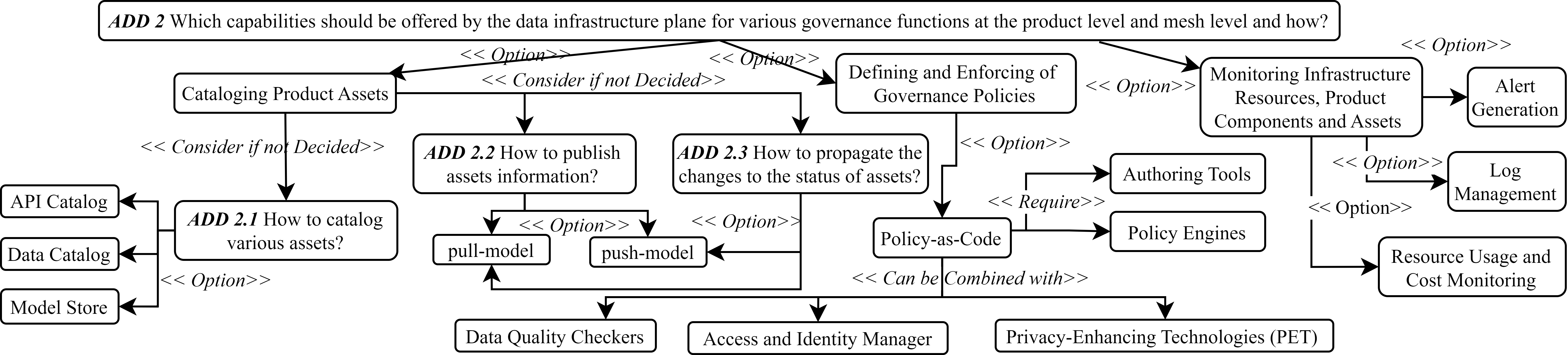}
 \caption{Decisions concerning Governance Support APIs}
 \label{fig:infra_goverment_component}
\end{figure*}

\textbf{Governance Support APIs.}
\Cref{fig:infra_goverment_component} shows the design decisions and their options to consider when designing the API to support the governance requirements of data products. 

A product must ensure that the data it consumes from upstream data products and the data it serves to the downstream products comply with the organization's various policies, standards, and conventions. For example, a data quality policy can state that the number of missing values for field X should be less than 1\% (completeness quality dimension), or the value of field Y should be in the format of Z (validity dimension). An example of a convention is using Parquet column-oriented files as the tables in the object stores. A privacy policy may dictate that Personal Identifiable Information (PII) should not be shared between some domains. Hence, a critical decision is how to empower product developers to enforce various policies on their data products. The potential options are to provide the tools (at the infrastructure plane) to author, validate, and execute policies for different data assets. While manually executing the policies is an option, our sources recommend support for their automated execution. In particular, they mentioned using the policy-as-code approach. The policy-as-code defines and manages policies (rules and conditions) through code, which enables versioning, auditing, and programmatically evaluating policies~\cite{das2023security}. The platform team needs to integrate policy enforcement mechanisms such as role-based access controllers, privacy-enhancing technologies~\cite{HEURIX20151} (e.g., data anonymization and encryption tools), and data quality checkers with policy-as-code tools to simplify implementing various policies (by domain and governance teams). 

The potential data product consumers need to be able to discover data products and find more information about them, including data models and access methods and policies. The infrastructure plane should have components allowing data product developers to publish the desired information about their products. A platform architect needs to decide what those components are and how they should be implemented and integrated to optimize the experience for product developers and consumers. According to our sources, the common components are data catalog, API catalog, and schema registry. A data catalog contains metadata of various data assets, including ML models and data sets. At the same time, a schema registry primarily stores the schemas of the data in transit or rest (e.g., the data exchanged between producers and consumers in event-driven applications or the data in a database). An API catalog stores various types of APIs. These components enable versioning, reusing, monitoring, and governing the corresponding assets. For example, an API catalog can provide the usage statistics of APIs and use role-based access control to restrict access to APIs.
The assets and their metadata can be pulled from a central component (e.g., a central data catalog) or pushed manually or automatically by a product product. While our sources use both options, in the data mesh, data products should be responsible for generating and publishing their metadata~\cite{dehghani2022data}. A critical decision is how to notify and propagate the changes to the key product assets to the consumers, for example, a change to a model or schema. The suggested options are pull-based (consumers checking for updates periodically) or push-based (sending updates to the subscribed consumers) models. The timely and safe propagation of changes can increase consumer satisfaction. 

Another area that platform architects must consider concerns the components that should be included to support monitoring data products and data mesh as a whole. Common options are resource monitoring, log management, and alert generation. These capabilities should also be isolated per domain or team. For example, the logs generated by a data product should only be visible to the product team.

\begin{figure*}[ht]
 \centering
 \includegraphics[width=0.80\textwidth]{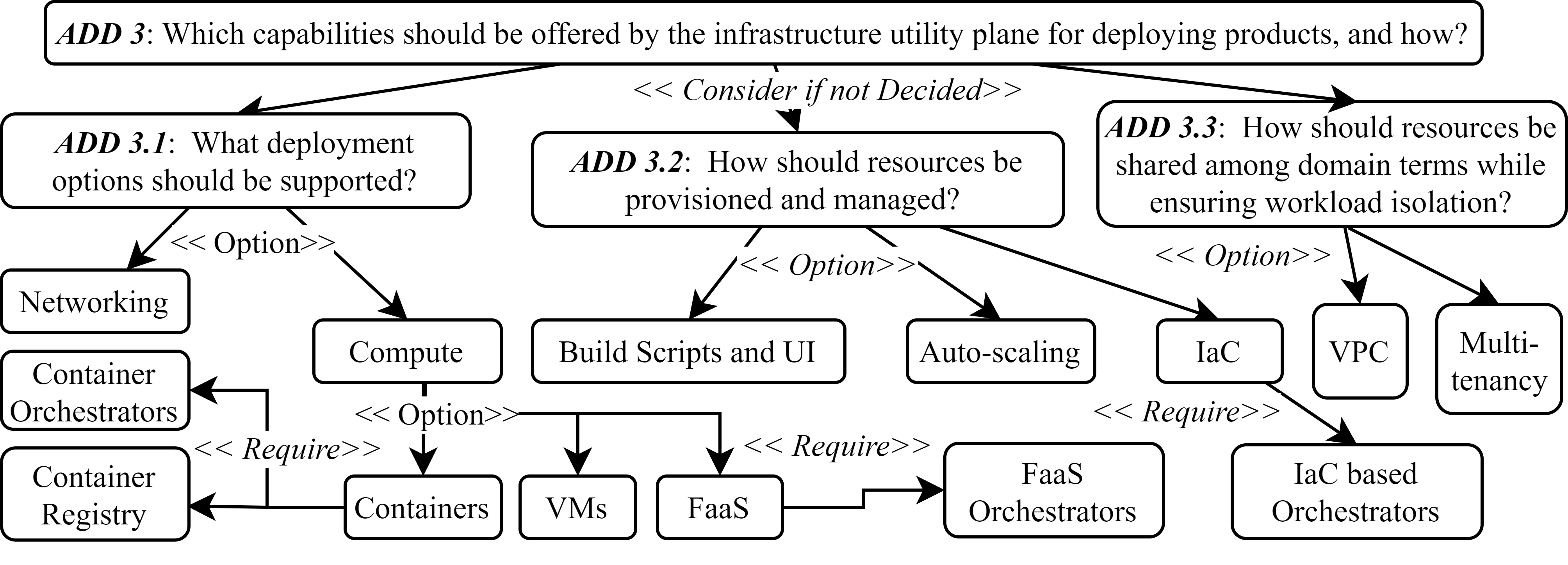}
 \caption{Decisions concerning Product Component Deployment APIs}
 \label{fig:infra_deployment_component}
\end{figure*}

\textbf{Product Component Deployment APIs.}
\Cref{fig:infra_deployment_component} shows the design decisions and their options to consider when supporting deploying product and platform components. 

The computing and networking resources are necessary for hosting the components of data products and the self-serve platform. A decision needs to be made about what infrastructure resources are used, how they are provisioned and managed securely and efficiently, and how their usage is measured. Our sources indicate that the deployment options of VMs, containers, FaaS (Function-as-a-Service) functions, and special-purpose hardware are used for executing product components. While different deployment options provide greater flexibility for product developers, without the tools for managing them at scale, it can incur considerable overhead for the platform team. 

Provisioning resources and deploying components can be done manually using the bash scripts, UIs, and command line tools or automatically using infrastructure as code (IaC) and CI-CD pipelines. For multi-container and FaaS applications, special orchestrators and hosting environments, such as OpenWhisk for FaaS and Kubernetes for containers, can deploy and manage those applications at scale. Using containers implies the platform should have a container registry to store and share images. IaC is the recommended option for deploying hybrid heterogeneous applications. With IaC, a platform engineer can define the desired configuration of the infrastructure in the source code and use an infrastructure orchestrator (IaC tool) to provision the infrastructure using the defined configuration~\cite{morris2020infrastructure,kumara2021s}. Typically, IaC scripts are stored in separate version-controlled repositories~\cite{kumara2021s}. 

Another key challenge for the platform team is to ensure the scalability and availability of the underlying infrastructure to maintain the quality of service (e.g., performance and cost) of product and platform components. The suggested option is to support auto-scaling for the infrastructure resources. However, as implementing auto-scaling for different platform components can pose a significant challenge for the platform team, a decision should be made on whether to build or buy an auto-scaling solution. 

A major issue in the shared infrastructure plane is to isolate data assets, compute resources, and performance at domain and product team levels so that an authorized domain/product team cannot access data sets and the resources are fairly allocated and accounted for. Our sources suggested that virtualization with multi-tenancy is the solution. Different levels of sharing and isolation may be supported. For example, each domain can be provided a virtual private cloud (VPC) with its resources and platform components, or domains can share a subset of resources and components. The complexity of implementing multi-tenancy may require considering building or buying the solution.

\begin{figure*}[ht]
 \centering
 \includegraphics[width=0.80\textwidth]{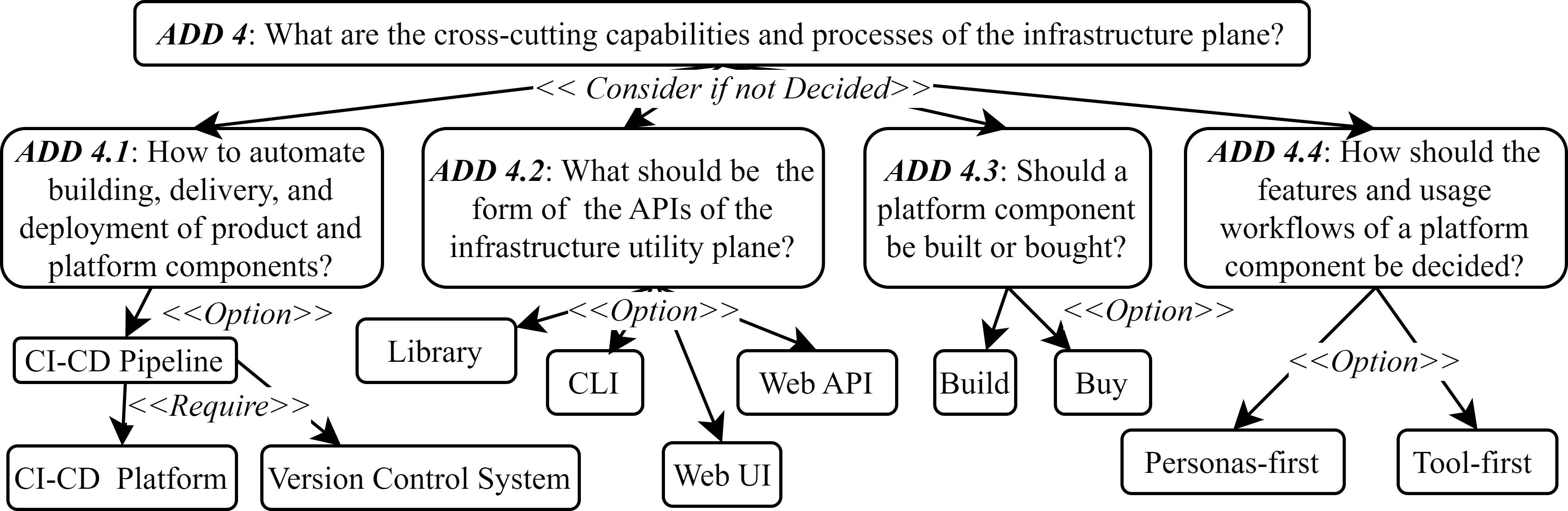}
 \caption{Common Decisions}
 \label{fig:infra_common_component}
\end{figure*}

\textbf{Common Decisions.}
\Cref{fig:infra_common_component} shows the design decisions and their options for cross-cutting capabilities and processes of the infrastructure plane. 

The components of data products and the self-serve platform are generally incrementally built, and the changes, such as bug fixes and new features, can be frequent. Product and platform developers must continuously test, build, integrate, deliver, and deploy those components. Our sources recommend using CI-CD pipelines to automate these processes. The infrastructure plane should include a CI-CD platform, version control systems for the source codes of the components and IaC scripts used for provisioning resources and deploying components, and storage for keeping multiple versions of the assets produced by the CI-CD pipelines and data products (e.g., model store and container registry).

The interfaces of the infrastructure plane should be designed to be used by the product and mesh experience planes. The common options in our sources are a library, Web API, Web UI, and a command line interface. A critical decision consideration concerns the interfaces that should be callable by an automation workflow (e.g., CI-CD pipeline, data pipeline, or ML pipeline). Otherwise, product developers could not automate the workflows of the data products (see \ref{sec-product-experience}).

A common decision related to any platform component is whether to build or buy the component (including open source). The buying option has two main sub-options: cloud and non-cloud. In the context of data mesh, a key design driver for selecting this option is the impact on the actors in the data mesh, particularly product developers, consumers, and platform developers. For example, there are cloud-based CI-CD solutions, and managing a CI-CD platform on-premise can be a significant overhead for the platform team, limiting their time and effort in addressing the needs for data domains. Another example is the implementation of auto-scaling and multi-tenancy for components, which is highly complex, and buying a component with the necessary support can be the appropriate solution. 

A related decision is what features a platform component should support. The two options are suggested: tool-first and personas-first. The tool-first approach selects a tool primarily based on its features with little or no consideration for the potential users of it in the data mesh. The second approach considers the characteristics and needs of the relevant actors first and chooses the tool that matches them. An example of our source is using a data processing tool that supports both Python-based API and SQL-based APIs since some domain teams have data engineers while others do not. Another example from our sources is to build a data catalog using a wiki or buying an existing feature-rich tool.


\begin{figure*}[ht]
 \centering
 \includegraphics[width=0.95\textwidth]{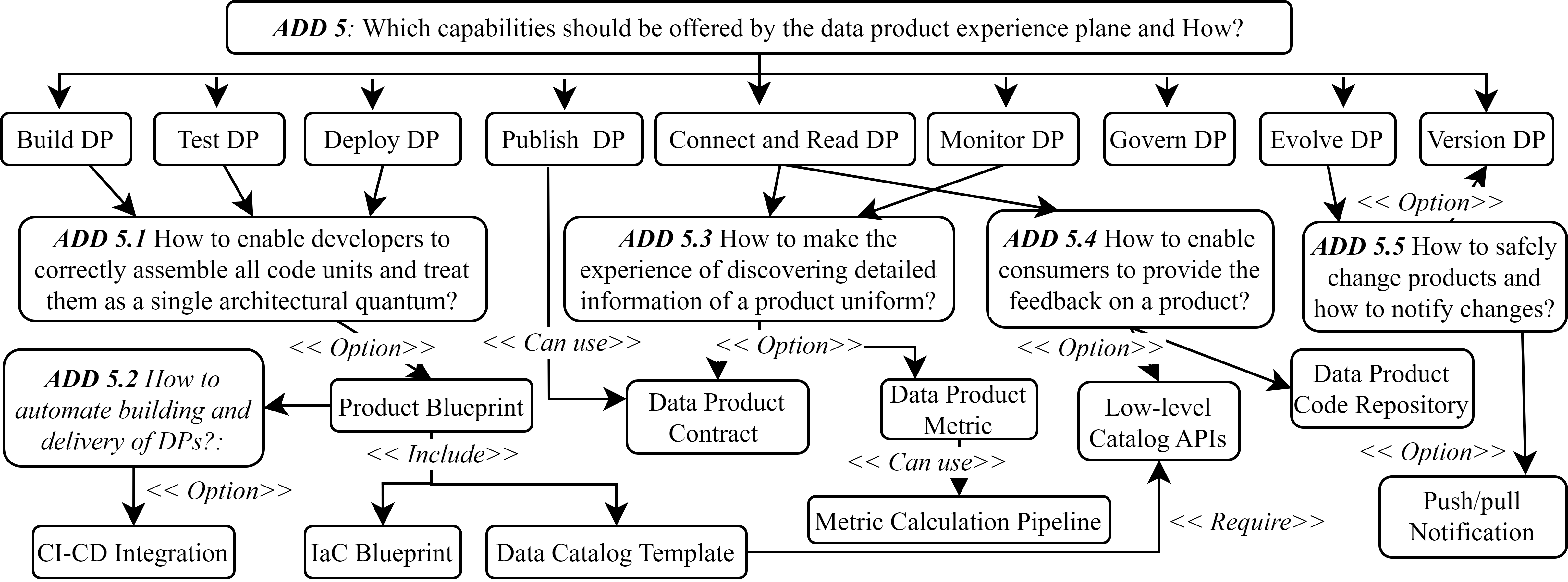}
 \caption{Decisions concerning the Data Product Experience Plane}
 \label{fig:data_product_developer}
\end{figure*}

\subsection{Decisions concerning the Data Product Experience Plane}
\label{sec-product-experience}
The data product experience plane aims to optimize the experience for data product developers and consumers by reducing cognitive load on them via providing high-level abstractions that consider all the architectural elements required for the product to function properly as a single cohesive unit (i.e., data product as an architectural quantum~\cite{dehghani2022data}). \Cref{fig:data_product_developer} shows the design decisions that a platform architect should consider when designing this platform plane. Within the scope of this paper, we do not consider the architectural decisions that a product developer, owner, or consumer should make when developing, operating, and consuming data products. Our focus is on architectural decisions that a platform team needs to make when designing the self-serve platform to support a product's lifecycle. 

The APIs of this plane aim to serve the product teams and consumers. Thus, a key decision is what capabilities the plane should offer to optimize the product team's experience in creating and managing data products. The options are to cover each step of the lifecycle of a data product: from developing to building, delivering, operating, and evolving it. A developer can use the APIs of the infrastructure plane to implement the individual elements of a data product, for example, data pipelines, data transformation scripts, data quality tests, and data privacy policies. The key decision to make here is how to enable developers to correctly assemble all code units and treat them as a single architectural quantum. According to our sources, some white literature~\cite{driessen2023promote,goedegebuure2023data}, and Zhamak~\cite{dehghani2022data}, the option is to use an abstraction of a product template or blueprint (a declarative definition of a product). A common use case for a product template is to act as a guide and documentation for a product~\cite{driessen2023promote}. However, to reduce the overhead of managing a product's life for a developer, a product template should support building, testing, and delivering a product automatically as a single unit. Our sources propose a product definition that can be integrated with a CI-CD process to build product components automatically, run all tests, deploy and configure components, and register the product. IaC blueprints (for deploying the product) and data catalog files (for registering the product) were two options mentioned in our sources. 

Regarding data product consumers, our sources identify two capabilities for the product experience plane: discovering detailed product information and providing feedback concerning a product. A consumer needs the information to assess the quality and health of a data product and to connect and consume the data assets from the output interfaces of the product. Common feedbacks are rating products, requesting access to data, adding new product features, and fixing bugs. The product experience plane needs to have mechanisms to support both capabilities. The suggested option is to provide a single unified entry point to all information using data product contracts and unified product-level metrics to ease the discovery and assessment of the data products. According to our sources, a data (product) contract between a consumer and a producer defines the terms and conditions on data usage, including the information about the data offered by a product (e.g., schemas and owners), access information (e.g., endpoints and required permissions), and agreements on data quality and service levels. An example of a unified metric is a trust metric that combines the ratings, likes, and number of activities related to a product's source code repository. Product developers need to understand the health of their data products in detail. Furthermore, a data product may use different input/output interfaces and data processing tools. Collecting custom metrics (e.g., metrics related to a pub-sub broker or a relational database) and aggregating them to create product-level metrics should be supported.  The implementation options for discovery and feedback features mentioned by our sources include the various catalog APIs from the infrastructure plane and a GitHub repository per product. The platform team must build metric calculation pipelines to extract the resource usage metrics and logs using the APIs provided by the infrastructure plane, derive product-specific metrics, and generate reports showing issues, their root causes, and resource costs. 

Another key challenge for the platform team is enabling product developers to make and manage product changes without undue overhead. Moreover, the changes to a data product's output interfaces (e.g., schema evolution) and health (e.g., unavailable for a particular duration or failure of data quality test) can affect consumers. Notifying changes in the status and quality of the data products can help improve the experience and satisfaction of the consumers. The solution suggested for accelerating the delivery of product changes is to use CI-CD automation. The versioning of data products can help consumers migrate to new products safely and allow rollback to previous versions as necessary. Similar to propagating changes at the asset level, the decision options for providing consumers with changes at the product level are push-based or pull-based models. 

The decisions related to the form of the plane's APIs and the processes of deciding on buying or building components are also relevant to the product experience plane. The previous section discussed them in detail.


\begin{figure*}[ht]
 \centering
 \includegraphics[width=0.70\textwidth]{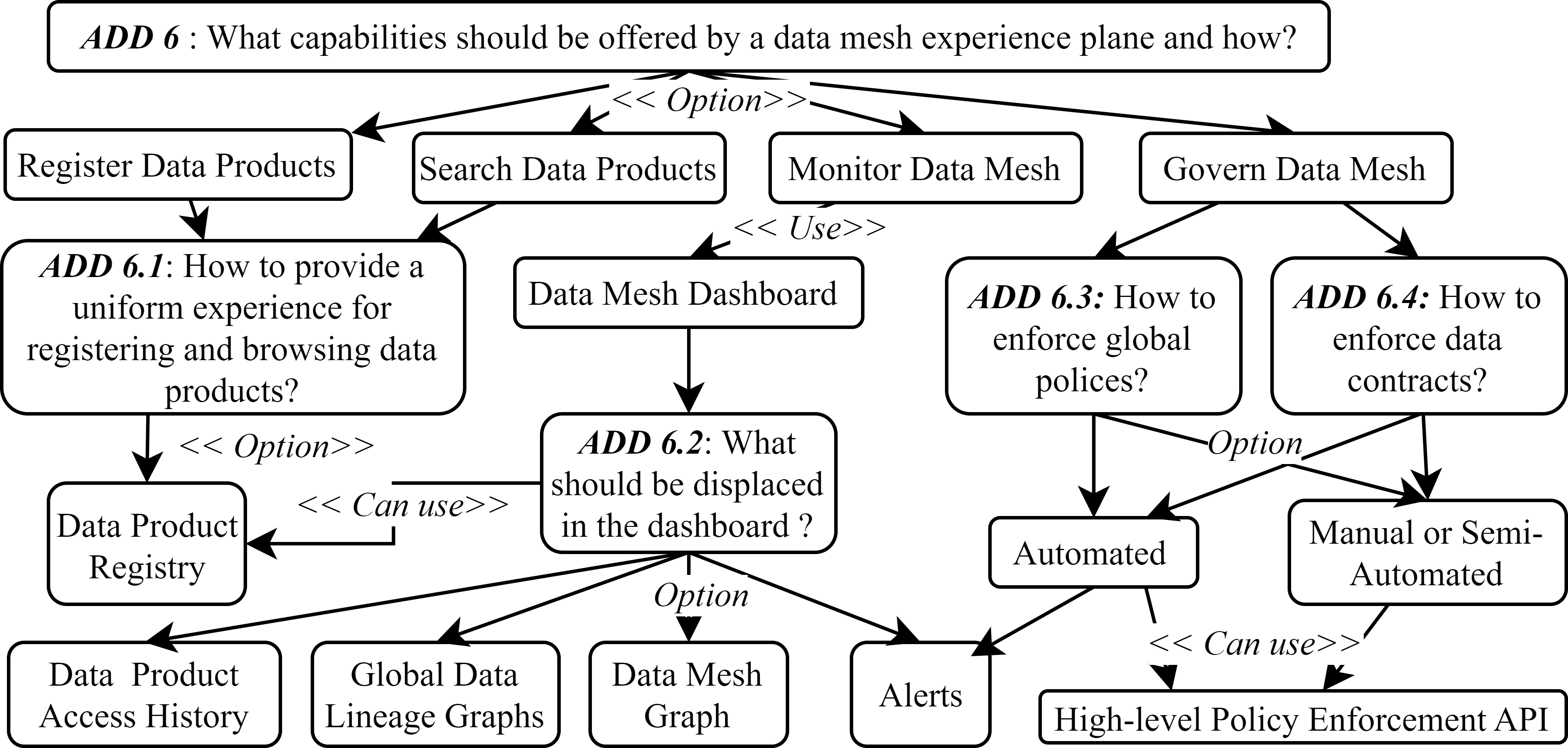}
 \caption{Decisions concerning the Data Mesh Experience Plane}
 \label{fig:meshdecision}
\end{figure*}

\subsection{Decisions concerning the Data Mesh Experience Plane}
\label{sec-mesh-experience}
\Cref{fig:meshdecision} shows the design decisions that the platform team should make when developing the data mesh experience plane. 

The product developers need the capability to add their products to the existing mesh of products or remove them from it. The options to support this are to use the data/API catalog APIs provided by the infrastructure plane or to provide a data product registry that hides the low-level catalog APIs by exposing the abstraction of a product. A consumer needs to be able to browse available data products in the mesh and search and select the appropriate data product. The search API can be implemented on top of the data product registry. According to our sources, to select a data product, a consumer may need more information, such as product history, data lineage (tracking of the history of the output data from a product to the sources), and syntactic and semantic similarities and variations in data. Thus, the mesh experience plane should provide these functions in its APIs. 

The governance team, including product owners, is the main user of the data mesh experience plane. The two essential capabilities are monitoring the data mesh and enforcing mesh-wide (global) policies. The governance team needs to be able to check the status of products and their connections to each other. Such information helps the team find top and unused products and assess the business value generated by products. A recommended option is to have a dashboard that depicts the data mesh graph with the necessary status. An example of a global policy is to change the data retention period.
Regarding enforcing policies, the governance team needs to be able to codify the policies (i.e., computational governance), apply them to the whole data mesh or a selected subset of products in a single operation, and collect the data necessary to ascertain the level of compliance with the policy. These APIs should be built using the APIs provided by the product experience and infrastructure plane. For example, a policy enforcement function in the mesh plane can execute the corresponding APIs of each product (provided by the product experience plane).

Another governance capability mentioned by several sources is monitoring data contract violations and automating the enforcement of data contracts. For example, the mesh experience plane can collect alerts related to data contracts and display them in a dashboard. It can monitor the changes to the schema of the outputs of data products, check the updated schemas' compatibility with the downstream consumers' schemas, and execute corrective actions if incompatibilities violate contacts between producers and consumers. An example of a corrective action is dropping a newly added field unused by a consumer or preventing the storage of incompatible data assets.


\begin{table}[t]
\centering
\caption{Perceived Usefulness and Ease of Use with Positive and Negative Comments}
\label{tab:SF-EI-Results}
\resizebox{0.99\linewidth}{!}{
\scriptsize
\begin{tabular}{|l|c|c|p{5cm}|} 
\hline
\textbf{Characteristic} & \textbf{\done} & \textbf{\wontfix} & \textbf{Prominent Comments}\\ 
\hline
\multirow{2}{*}{Perceived Ease of Use} & \multirow{2}{*}{4} & \multirow{2}{*}{3} & p4: ``Well, I think, well, here's how I would suggest... So if you were to provide this to a client, their head would spin. Yeah. Because it has too many choices...''\\
& & & p5: ``So to me with all the context, it's very easy to understand. Yeah, with the context. It's elegant because it has, you tie it to theory...''\\
\hline
\multirow{3}{*}{Perceived Usefulness} & \multirow{3}{*}{4} & \multirow{3}{*}{2} & p2: ``..., there are a lot of decisions that need to be taken when designing a data architecture, of course. So these mappings kind of help out in sorting them out''\\
& & & p5: ``This is a playbook and a cookbook to help you do that, that to me is powerful from a repeatability standpoint. It takes the theory and puts it into practice.''\\
& & & p6: ``But overall- Maybe not in the charts, ...if you build some kind of interactive version of this that people can use...''\\
\hline
\end{tabular}
}
\end{table}

\section{Interview-based Refinement and Validation}
\label{sec:validation}
This section describes the assessment conducted through expert interviews concerning the completeness and usefulness of our previous analyses. 

\subsection{Completeness Evaluation}
This section provides a qualitative analysis of the interviews, focusing on changes made to the original framework. 

\textbf{Inclusion and Exclusion of Decisions and Options.}
Participants generally agreed with most of the decisions and options identified from the gray literature, suggesting minimal changes. For example, p5 commented on the product experience plane options, \textit{"I believe you have captured nearly everything that we have."}.
P4 strongly agreed with the options in the infrastructure plane, \textit{"...A lot of the capability obviously is available with some of the cloud vendors. But I think you cover all the bases. I can't think of anything that you've omitted. So this looks really good"}. 

Experts helped to identify technology-specific options and replace them with technology-agnostic decision options. For example, we removed the options for organizing a self-serve platform into zones specific to a cloud provider. The options that violate the principles of data mesh were also identified based on the experts' feedback. For example, multiple practitioners stressed that a central master data management strategy would lead to a monolithic data mesh implementation, preferring domains to manage their master data. The expert also suggested new options we overlooked in the gray literature or those that improve compliance with the data mesh principles. For example, our initial framework did not include the mechanisms for managing product changes. After a suggestion from an expert, we checked our literature sources and discovered it was an important decision.
Another example is the persona-centric approach option for the decision to build or buy a platform component/tool, which was also advocated by Zhamak~\cite{dehghani2022data}. The expert feedback was useful for extending some options. For example, the participants suggested that the platform should support diverse options for adding datasets to a data catalog, a low-code data transformation tool that caters to non-engineering users, and multiple data ingestion patterns. 

\textbf{Reorganization of Decisions and Options.}
The feedback from experts led to changes to the organization of the decisions and their options. First, we removed a major decision category (the user interface of a self-serve platform) and assigned decisions and options in that category to the three existing planes. The feedback from the experts revealed that such a decision category was unnecessary and violated the separation of concerns prompted by the three-plane logical architecture. For example, p5 stated, \textit{"... plane architecture is general enough, and it allows you to be flexible and have to capture all the different pieces ... it covers 99\% of the use cases." }. Second, we moved decisions and options between planes and merged and split options. For example, we separated the data catalog tool (the infrastructure plane) from the product registration function (mesh supervision), moved the schema registry to the infrastructure plane, and used specific capabilities such as product registration and discovery instead of the generic capabilities metadata management. Finally, we renamed some options based on the suggestions from the experts. For example, we changed the option \textit{interconnectivity} into \textit{networking}.

\subsection{Usefulness Evaluation}
\Cref{tab:SF-EI-Results} shows the summary of the evidence for \textit{Perceived Usefulness} and \textit{Perceived Ease of Use} of our decision framework. Overall, the participants acknowledged the importance of having a comprehensive catalog of all decisions, their options, and guidelines for selecting them. However, the complexity induced by having many selection options can make using the framework challenging. They also suggested techniques for improving the framework's usefulness: building an interactive tool, using defaults for decisions based on the information about the target use case, and relating the decision options to the tools available for realizing them. 

\section{Threats to Validity}
\label{sec:threats}
Our study can exhibit the common types of potential threats for qualitative research. The typical threats to a gray literature review include bias in selecting and interpreting articles and the relevance of the selected articles. To ensure the selection of the most relevant papers, we used multiple search queries, reference lists and backlinks from the sources, explicit inclusion and exclusion criteria, quality assessment criteria, and inter-rater reliability assessment. As discussed in \Cref{sec:design}, we followed standard practices in qualitative research for coding gray literature sources to limit observer and interpretation bias. However, we might not have eliminated the researcher bias, as usual in qualitative research. 

Potential threats in the semi-structured interview study include the bias in selecting participants, the relevance of participants, the limited number of participants, and the data collection and analysis. To limit this selection bias, we use an open data mesh learning community as the main source of finding participants. Before contacting potential candidates, we carefully examined the participants' profiles to ensure they have relevant expertise for the study. The selected participants were from organizations with distinct sizes from multiple domains. They also had diverse roles and experiences in relevant topics such as data engineering, cloud engineering, consultancy (data mesh), and data mesh transition.
Regarding the number of participants, we believe the sample size was sufficient to assess and improve the findings from the literature. The interview questions and protocol can also threaten the construct validity. To mitigate this threat, we followed a standard protocol and identified the questions from the related literature (see \Cref{sec:design}). We also pre-tested the interview protocol with a researcher working on data mesh and data market. We followed standard practices for the thematic data analysis for analyzing and interpreting interview transcripts.

\section{Conclusions and Future Work}
\label{sec:conclusions}
This study synthesizes architectural design decisions (ADDs) and options for self-serve platforms in data meshes from a systematic review of the relevant gray literature and a semi-structured interview with six experts. We identified six main ADDs mapped to the three planes of the logical architecture of a self-serve platform. We also identified potential decision options for each ADD and their impact on the experience of the data mesh stakeholders. We believe our findings can help organizations systemically (re-)design their self-serve platforms to accelerate their data mesh transition. Our findings can also be beneficial to researchers in identifying key design and implementation issues in a self-serve platform. 

We have also been developing similar architectural decision frameworks for data products and federated data governance leveraging gray literature and semi-structured interviews. We aim to build a decision support system that allows practitioners to interactively explore ADDs and design options, find the tools for realizing the selected options, and provide feedback on the selected options and tools. Finally, an important research direction is to further refine and extend our findings through more interviews with experts, continuous analysis of rapidly growing gray literature on data mesh, and action research studies in diverse organizational settings.

\section*{Acknowledgment}
This research has received funding from the European Union's Horizon research and innovation program under the grant agreement No 101097036 (ONCOSCREEN).
\bibliographystyle{IEEEtran}
\bibliography{main}

\end{document}